\documentstyle[12pt]{article}

\newcommand{\EQ}{\begin{equation}}
\newcommand{\EN}{\end{equation}}
\def\aprle{\buildrel < \over {_{\sim}}}
\def\aprge{\buildrel > \over {_{\sim}}}
\begin{document}
\topmargin 0pt
\oddsidemargin=-0.4truecm
\evensidemargin=-0.4truecm
\renewcommand{\thefootnote}{\fnsymbol{footnote}}
\newpage
\setcounter{page}{0}
\begin{titlepage}
%\vspace{0.5cm}
\begin{flushright}
FTUV/94--26\\
15A-40561-INT94-13\\
\end{flushright}
%\vspace{0.2cm}
\begin{center}
{\large NEUTRINO MAGNETIC MOMENTS AND THE SOLAR NEUTRINO PROBLEM}
\footnote{Talk given at the 6th International Symposium ``Neutrino
Telescopes'', Venice, February 22--24, 1994} \\
%\vspace{0.4cm}
\vspace{0.5cm}
{\large E.Kh. Akhmedov}
\footnote{On leave from NRC ``Kurchatov Institute'', Moscow 123182, Russia}
\vspace{0.5cm}\\
{\em Institute of Nuclear Theory, Henderson Hall, HN-12,\\
University of Washington, Seattle, WA 98195}\\
%\vspace{0.3cm}
{\em and}\\
%\vspace{0.3cm}
{\em Instituto de Fisica Corpuscular (IFIC--CSIC)\\
Departamento de Fisica Teorica, Universitat de Valencia\\
Dr. Moliner 50, 46100 Burjassot (Valencia), Spain}\\
\vspace{0.5cm}
\end{center}
%%\vspace{0.4cm}
%\vspace*{-0.2cm}
\begin{abstract}
Present status of the neutrino magnetic moment solutions of the solar
neutrino problem is reviewed. In particular, we discuss a possibility of
reconciling different degrees of suppression and time variation of the signal
(or lack of such a variation) observed in different solar neutrino
experiments. It is shown that the resonant spin--flavor precession of
neutrinos due to the interaction of their transitions magnetic moments
with solar magnetic field can account for all the available solar neutrino
data. For not too small neutrino mixing angles ($\sin 2\theta_0 \aprge 0.2$)
the combined effect of the resonant spin--flavor precession and neutrino
oscillations can result in an observable flux of solar $\bar{\nu}_{e}$'s.
\end{abstract}
\end{titlepage}
%\end{document}
\vspace{2cm}
\vspace{.5cm}
%\end{document}
\renewcommand{\thefootnote}{\arabic{footnote}}
\setcounter{footnote}{0}
\newpage
\section{Introduction}
The solar neutrino problem, i.e. the deficiency of the observed flux of
solar neutrinos
as compared to the predictions of the standard solar model,
remains one of the major unresolved puzzles of modern physics and
astrophysics. Although the astrophysical solution of the problem is not yet
completely ruled out, it is very unlikely to be the true reason
of the discrepancy provided all the experimental data (and in particular,
the results of the Homestake experiment for the whole period of its
operation) are taken seriously. It is for this reason that the
particle--physics solutions to the problem are currently considered to be
more favorable
\cite{BHL,B,Sm1}.

There are several possible neutrino--physics solutions of the solar
neutrino problem, the most popular one being resonant neutrino oscillations
in the matter of the sun (the MSW effect \cite{MSW}). In my talk I will
concentrate, however, on another type of solutions related to possible
existence of large magnetic or transition magnetic moments of neutrinos.
In this case neutrino spin precession \cite{C,VVO} or spin--flavor
precession \cite{VVO,Akhm1,LM} can occur in the magnetic field of the sun,
converting a fraction of solar $\nu_{eL}$ into $\nu_{eR}$ or into
$\nu_{\mu R}$, $\nu_{\tau R}$, $\bar{\nu}_{\mu R}$ or $\bar{\nu}_{\tau R}$.
Although $\bar{\nu}_{\mu R}$ and $\bar{\nu}_{\tau R}$ are not sterile, they
cannot be observed in Homestake, SAGE and
GALLEX experiments and can only be detected with a small cross section in
the Kamiokande experiment. Spin--flavor precession of neutrinos can be
resonantly enhanced in the matter of the sun \cite{Akhm1,LM}, in direct
analogy with the MSW effect. Neutrino spin precession and resonant
spin--flavor precession (RSFP) can account for both the
deficiency of solar neutrinos and time variations of the solar neutrino
flux in anticorrelation with solar activity for which there are some
indications in the Homestake data. This comes about because the toroidal
magnetic field of the sun is strongest in the periods of active sun.

Some remarks about time structure of the Homestake data are in order.
The data compares better with an assumption of a time--dependent signal
than with that of a constant one, hinting to an anticorrelation with solar
activity. The existing analyses of the data using different statistical
methods gave fairly big values of the correlation coefficient between the
data and sun--spot number \cite{BP,BSSS,K,FV,NM}. These analyses, however,
were performed before 1990 and so did not take into account more recent
runs 109-126. These runs
do not show a tendency to vary in time, similarly to runs 19--59. A recent
analysis of Stanev \cite {St} which updated the one of ref. \cite{BSSS}
included the runs 109-126 and showed that this results in the correlation
coefficient being decreased by an order of magnitude as compared to the
previously obtained one, but the correlation probability is still large:
confidence level of the correlation with the sunspot number $s$ is 0.96
instead of 0.996, and that of the correlation with $s|z|$ where $z$ is the
latitude of the line of sight is 0.99 instead of 0.9993.  The correlation
with the 22-yr cycle is even better than the correlation with the 11-yr
one. Therefore, the possibility that the solar neutrino flux
anticorrelates with solar activity still persists and deserves further
study.

At the same time, the Kamiokande group did not observe any time variation
of the solar neutrino signal in their experiment, which allowed them to put
an upper limit on the possible time variation, $\Delta Q/Q < 30\%$ at 90\%
c.l.. Therefore a question naturally arises as to whether one can reconcile
a strong time variation in the Homestake experiment with a small (or no)
time variation in Kamiokande. Recently, it has been shown \cite{ALP1} (see
also \cite{MN2,Pu,Kr} that
the RSFP scenario is capable of accounting for all the existing solar
neutrino data, including their time structure or lack of such a structure.
In particular, it can explain mild suppression of
the flux in the gallium experiments and naturally reconcile strong
time variations of the signal observed in the Homestake experiment with
small time variations allowed by the Kamiokande data. Let me
now briefly describe how this works.
\section{GALLEX and SAGE}
Although the gallium solar neutrino experiments SAGE and GALLEX have been
operating for too short a time and so are unable to confirm or
disprove 11-yr variations of the signal, they still provide us with
an information which is relevant for the magnetic moment scenarios.
The point is that most of the data have been taken during the period
of high solar activity. One could therefore expect a strong suppression of
the signal in the gallium experiments, which has not been observed. This
disfavors the ordinary spin precession scenario since it is neutrino-energy
independent  and so predicts the same degree of suppression and time
variation of the signal in all the solar neutrino experiments.  At the
same time, the resonant spin--flavor precession (RSFP) is strongly energy
dependent and so naturally results in different suppressions and
time variations in different experiments. In particular, the $pp$ neutrinos
which are expected to give the major contribution to the signal in the
gallium experiments,
have low energies and so should encounter the RSFP resonance at high
densities, somewhere in the radiation zone or in the core of the sun (since
the resonant density is inversely proportional to neutrino energy). We know
that the magnetic field does exist and may be quite strong in the convective
zone of the sun ($0.7R_\odot\leq r \leq R_\odot$). However, it is not clear
if strong enough magnetic field can exist deeper in the sun, i.e. in the
radiation zone or in solar core. If the inner magnetic field of the sun is
week, the RSFP will not be efficient there and the $pp$ neutrinos will
leave the sun intact, in accordance with the observations of GALLEX and SAGE.
One can turn the argument around and ask the following question: If we
believe in the RSFP mechanism, what is the maximal allowed inner magnetic
field which is not in conflict with the gallium experiment? The answer turns
out to be $(B_i)_{max}\approx 3\times 10^6$ G assuming the neutrino
transition magnetic moment $\mu=10^{-11}\mu_B$ \cite{ALP1}.
\section{Reconciling Homestake and Kamiokande data}
It is more difficult to explain how one can reconcile strong time
dependence of the signal observed in the Homestake experiment with no or
very little time variation of the Kamiokande data. The key points here are
that \cite{Akhm2,BMR,OS,ALP1}

(1) The two experiments are sensitive to slightly different parts of the
solar neutrino spectrum: Homestake is sensitive to both energetic ${}^8$B
neutrinos and medium--energy ${}^7$Be and $pep$ neutrinos, whereas the
Kamiokande experiment is only sensitive to the high--energy part of ${}^8$B
neutrinos ($E>7.5$ MeV);

(2) For Majorana neutrinos, the RSFP converts left--handed $\nu_{e}$ into
right--handed $\bar{\nu}_{\mu}$ (or $\bar{\nu}_{\tau}$) which are sterile for
the Homestake experiment (since their energy is less than the muon or tauon
mass) but do
contribute to the event rate in the Kamiokande experiment through their
neutral--current interaction with electrons. Although the $\bar{\nu}_{\mu}e$
cross section is smaller than the $\nu_{e}e$ one, it is non-negligible,
which reduces the amplitude of the time variation of the signal in the
Kamiokande experiment.

It turns out that the above two points are enough to account for the
differences in the time dependences of the signals in the Homestake and
Kamiokande experiments. The calculated event rates in the Homestake and
Kamiokande experiments decrease with increasing convective zone magnetic
field until they reach their minima, and then start to increase. The
minimum of the Kamiokande signal is situated at a lower magnetic field then
the one of the Homestake signal due to the energy dependence of the RSFP
and the above point (1). Also, it is shallower than the minimum of the
Homestake signal due to the point (2). For these reasons, for a certain
range of variation of the solar magnetic field $B_{\bot}$ the Homestake
signal can decrease significantly with increasing $B_{\bot}$ whereas
the Kamiokande signal is near its minimum and therefore does not change
much \cite{BMR,OS,ALP1}.
\section{Fitting the data}
In a recent paper \cite{ALP1} all the available solar neutrino data have
been analyzed in the framework of the RSFP disregarding neutrino mass
mixing. However, in the general case one should include neutrino mixing
effects as well. The motivation for that is as follows:

(1) RSFP requires non-vanishing flavor-off-diagonal neutrino magnetic
moments, i.e. implies lepton flavor non-conservation. Therefore neutrino
oscillations $must$ also take place. In general one should therefore
consider the RSFP and neutrino oscillations (including the MSW effect)
jointly. The results of ref. \cite{ALP1} are only valid in the small mixing
angle limit.

(2) It has been shown in \cite{ALP1} that all the existing solar neutrino
data can be fitted within the RSFP scenario for certain model magnetic field
profiles and certain values of neutrino parameters $\mu$ and $\Delta m^2$.
It would be interesting to see how the neutrino mixing modifies these
results.

(3) In ref. \cite{Akhm3} it has been suggested that the combined action of
the RSFP and MSW effect in the convective zone of the sun can relax the lower
limit on the product $\mu B_{\bot}$ of neutrino magnetic moment and
solar magnetic field required to account for the data. The main idea was
that the MSW effect can assist the RSFP to cause the time variations of the
neutrino flux by improving the adiabaticity of the RSFP (this can occur
when the RSFP and MSW resonances overlap). It would be interesting to
confront this idea with the new experimental data.

Combined action of the RSFP and the MSW effect on solar neutrinos has been
considered in a number of papers \cite{LM,Akhm4,MN1,BalHL,Akhm5}.
However, the data of the gallium experiment were not available that time.
We therefore re-analyzed all the available
solar neutrino data in the framework of the RSFP scenario taking into
account possible neutrino mixing and oscillations effects
\cite{ALP2}. It was assumed that the $\nu_{e}$ -- $\nu_{\mu}$ mixing is
generated by a Majorana neutrino mass term, and that there exists a
$\nu_{eL}$ -- $\bar{\nu}_{\mu R}$ transition magnetic moment $\mu$. The
evolution equation for a system of two Majorana neutrinos and their
antiparticles in the flavor basis is
\EQ
i\frac{d}{dt}\left(\begin{array}{l}
   \nu_{eL}\\
   \bar{\nu}_{eR}\\
   \nu_{\mu L}\\
   \bar{\nu}_{\mu R}
\end{array}\right )
{}~=~\left (
\begin{array}{cccc}
   N_1-c_{2}\delta  & 0 & s_{2}\delta & \mu B_{\bot}\\
  0 & -N_1-c_{2}\delta & -\mu B_{\bot} & s_{2}\delta\\
  s_{2}\delta & -\mu B_{\bot} & N_2+c_{2}\delta & 0\\
  \mu B_{\bot} & s_{2}\delta & 0 & -N_2+c_{2}\delta
\end{array}\right )\left(\begin{array}{l}
   \nu_{eL}\\
   \bar{\nu}_{eR}\\
   \nu_{\mu L}\\
   \bar{\nu}_{\mu R}
\end{array}\right)
\EN
Here $B_{\bot}(t)$ is the transverse magnetic field,
$$N_1\equiv \sqrt{2}G_{F}(n_{e}-n_{n}/2),~~N_2\equiv \sqrt{2}G_{F}(-n_{n}/2),
{}~~\delta\equiv \Delta m^{2}/4E,$$
\EQ
s_{2}\equiv\sin 2\theta_{0},~~c_{2}\equiv \cos 2\theta_{0},
\EN
where $G_{F}$ is the Fermi constant, $n_{e}$ and $n_{n}$ are the electron
and neutron number densities, the rest of the notation being obvious.
The zeros in the effective Hamiltonian in eq. (1) are related to the fact
that diagonal magnetic moments of Majorana neutrinos are precluded by $CPT$
invariance.

We have calculated the neutrino signals in the chlorine, gallium and
Kamiokande experiments using ten different model magnetic field profiles
(see \cite{ALP2} for more details). The results of our analysis are briefly
summarized below.

(1) For small mixing angles, $\sin 2\theta_0\aprle$0.1, the results of
our previous study \cite{ALP1} are only slightly modified.

(2) For moderate mixing angles, $\sin 2\theta_0\aprge$0.2, some of the
magnetic field profiles which proved to give good fit of the data for
vanishing
$\theta_0$, no longer work: they result in too strong a suppression of the
signal in the gallium experiments since the adiabaticity of the MSW effect
for the low--energy $pp$ neutrinos gets too good. Reasonable fit can still
be achieved for very large mixing angles, $\sin 2\theta_0\approx 1$, but in
this case a large flux of electron antineutrinos would be produced in
contradiction with an upper limit derived from the Kamiokande and LSD data
\cite{BFMM,LSD} (see below, point (5)).

(3) Possible way out of this situation is to use the model magnetic field
profiles with their maximum being shifted towards the outer regions of the
convective zone. This would require lower values of of $\Delta m^2$ for the
RSFP to be efficient, which in turn would decrease the adiabaticity of the
MSW effect, and the flux of the $pp$ neutrinos will be essentially
unsuppressed. We have tried three such new magnetic field configurations
and they produced good fit of all the data.

(4) Typical values of the neutrino parameters required to account for the
data are $\Delta m^2 \simeq (10^{-8}$--$10^{-7})$ eV$^2$, $\sin 2\theta_0
\aprle$ 0.2--0.4, depending on the magnetic field configuration; for neutrino
transition magnetic moment $\mu=10^{-11}\mu_B$ the maximum magnetic field in
the solar convective zone should vary in time in the range (15--30) kG.

(5) As have been noticed above (points (2) and (3)), some magnetic field
configurations which used to give a good fit to the data for vanishing
$\theta_0$, no longer do so for not too small mixing angles and, conversely,
some other profiles which failed to reproduce the data for $\theta_0=0$
do give a good fit for moderate $\theta_0$. This is, in fact, a rather
unpleasant situation: whether or not a given magnetic field profile fits
the data depends on the neutrino mixing angle which is unknown. Possible
way out is to look for the $\bar{\nu}_{eR}$ signal from the sun. The point
is that if neutrinos experience the RSFP in the sun and also have mass
mixing, a flux of electron antineutrinos can be produced which is in
principle detectable in the SNO, Super--Kamiokande and Borexino experiments
even in the case moderate neutrino mixing angles
\cite{LM,Akhm4,Akhm5,RBLBPP,BL1}.
The main mechanism of the $\bar{\nu}_{eR}$ production is $\nu_{eL}\rightarrow
\bar{\nu}_{\mu R}\rightarrow \bar{\nu}_{eR}$, where the first transition
is due to the RSFP in the sun and the second one is due to the vacuum
oscillations of antineutrinos on their way between the sun and the earth.
The salient feature of this flux is that it should vary in time in $direct$
correlation with solar activity. The detection of the solar $\bar{\nu}_{eR}$
flux would be a signature of the combined effect of the RSFP and neutrino
oscillations. It could allow one to discriminate between small mixing angle
and moderate mixing angle solutions.

The $\bar{\nu}_{eR}$ flux can be significantly
enhanced if the solar magnetic field changes its direction along the
neutrino trajectory \cite{APS1,APS2,BL2}. In this case one can have a
detectable $\bar{\nu}_{eR}$ flux even if the neutrino magnetic moment is
too small or the solar magnetic field is too weak to account for the solar
neutrino problem \cite{BL2,AS}.

To summarize, the RSFP is a viable scenario which is capable of accounting
for all the presently existing solar neutrino data, including their time
structure (or lack of such a structure). It also gives very specific
predictions for the forthcoming solar neutrino experiments, such as strong
time dependence of the ${}^7$Be neutrino flux, absence of a suppression and
time variation in the neutral--current events at SNO, and an observable
flux of solar $\bar{\nu}_{eR}$'s for moderate neutrino mixing angles.
\section*{Acknowledgements}
The hospitality of the National Institute for Nuclear Theory at the
University of Washington where this work was completed is gratefully
acknowledged. This work has been supported by the sabbatical grant from
Spanish Ministry of Education and Science and by the U.S. Department of
Energy under grant DE-FG06-90ER40561.
%\newpage

\end{document}